\PassOptionsToPackage{hyphens}{url}
\documentclass[aps,pre,reprint,superscriptaddress,nofootinbib]{revtex4-2}
\usepackage[T1]{fontenc}
\usepackage[utf8]{inputenc}
\usepackage{amsmath,amssymb,bm}
\usepackage{graphicx}
\usepackage{booktabs}
\usepackage{array}
\usepackage{xcolor}
\usepackage{url}
\newcommand{\avg}[1]{\langle #1 \rangle}
\newcommand{\XiEff}{\Xi_{\rm eff}}
\newcommand{\XiZero}{\Xi_0}
\newcommand{\Bin}{B_{\rm in}}
\newcommand{\Dz}{D_z}
\newcommand{\Dx}{D_x}
\newcommand{\Peff}{P_{\rm eff}}
\newcommand{\btr}{\beta_{\rm tr}}
\newcommand{\de}{d_e}
\newcommand{\tausp}{\tau_{\rm sp}}
\newcommand{\mupar}{\mu_\parallel}
\newcommand{\Jm}{\mathbf J_M}
\newcommand{\Jfree}{\mathbf J_{\rm free}}
\newcommand{\Mfield}{\mathbf M}
\newcommand{\safeincludegraphics}[2][]{%
  \IfFileExists{#2}{\includegraphics[#1]{#2}}{\fbox{\parbox{0.86\linewidth}{\centering Missing figure file: \texttt{\detokenize{#2}}}}}%
}

\begin{document}

\title{Relativistic Scaling and Magnetization-Current Feedback in Stern–Gerlach-Modified Pair-Plasma Reconnection: SpinPIC2D Validation and Nonlinear Regimes}

\author{K. Nykyri}
\affiliation{Physical Sciences Department and Centre for Space and Atmospheric Research, Embry--Riddle Aeronautical University, Daytona Beach, Florida 32114, USA}

\begin{abstract}
We investigate the relativistic scaling and electromagnetic feedback of
Stern--Gerlach (SG) force driven spin transport in pair-plasma reconnection with
SpinPIC2D. The model advances relativistic proper momentum and magnetic BMT
spin precession, applies the SG force, deposits spin magnetization, and
includes $\mathbf J_M=\nabla\times\mathbf M$ in Ampere's law. In a weak-seed
scan at fixed $\gamma_{\rm tr}=2$, the normalized global magnetic flux-growth 
remains near the classical control-run for $\Xi\le0.1$, is $0.016$ at $\Xi=0.4$, and reaches
approximately $0.12$ and $0.25$ at $\Xi=0.7$ and 1, respectively over
$3\le t/\tau_{\rm sp}\le7$. Because $\partial/\partial y=0$ in the 2.5-D
geometry, the direct $y$-directed SG term vanishes and the enhancement is
indirect: sheet-normal SG sorting restructures branch-resolved electron and positron velocity distributions which 
results in changes in pressure moments and generates a layered magnetization current. Along a
fixed-$\chi_{\rm sim}$ family, increasing $\gamma_{\rm tr}$ reduces $\Xi$ and
suppresses branch sorting, whereas the matched-control flux-growth enhancement
remains positive for $\gamma_{\rm tr}=2,3,5$. Retuning $\chi_{\rm sim}$ to hold
$\Xi=1$ does not preserve nonlinear similarity: both the coupling and
$J_{M,y}$ increase with $\gamma_{\rm tr}$, and the $\gamma_{\rm tr}=5$ case
develops a multi-X-line state. Thus $\Xi$ orders the onset of SG-modified reconnection, while the
nonlinear response also depends on the absolute spin coupling and
magnetization-current amplitude.
\end{abstract}

\maketitle

\section{Introduction}

Magnetic reconnection converts magnetic energy into plasma flow, heat, and nonthermal particles through localized regions where magnetic topology changes. Relativistic pair-plasma reconnection is relevant to compact-object coronae, pulsar-wind nebulae, relativistic jets, and magnetar magnetospheres~\cite{Kagan2015,Sironi2014,Uzdensky2011}. In such systems a central question is whether quantum spin physics remains a negligible microscopic correction or can organize macroscopic current-sheet dynamics. Semiclassical spin-kinetic theory predicts both Stern--Gerlach (SG) forces and magnetization currents~\cite{Marklund2007,BrodinMarklund2007,Zamanian2010}, but these terms are absent from conventional relativistic reconnection PIC studies. The companion Letter~\cite{NykyriPRLCompanion} introduces a compact SG control parameter and reports magnetic-moment branch sorting, pressure-moment restructuring, magnetization-current formation, and enhanced reconnected-flux growth at fixed $\gamma_{\rm tr}=2$. The present paper documents the numerical implementation and validation, extends the comparison to $\gamma_{\rm tr}=2,3,5$, and examines the separate scaling of branch sorting and magnetization-current feedback. The production runs are deliberately focused on the seeded onset and early nonlinear response, where the matched cases can be compared before their island-merger and saturation histories diverge: the normalized rate is measured over $3\le t/\tau_{\rm sp}\le7$, and the morphology is followed through $t=12\tau_{\rm sp}$. By the end of this interval the sorting, pressure response, magnetization current, and flux-growth enhancement are already established; the long classical control in Fig.~S9 of the Supplemental Material remains stable, confirming that classical PIC implementation is energy conserving.

\begin{figure}[!t]
\centering
\safeincludegraphics[width=\columnwidth]{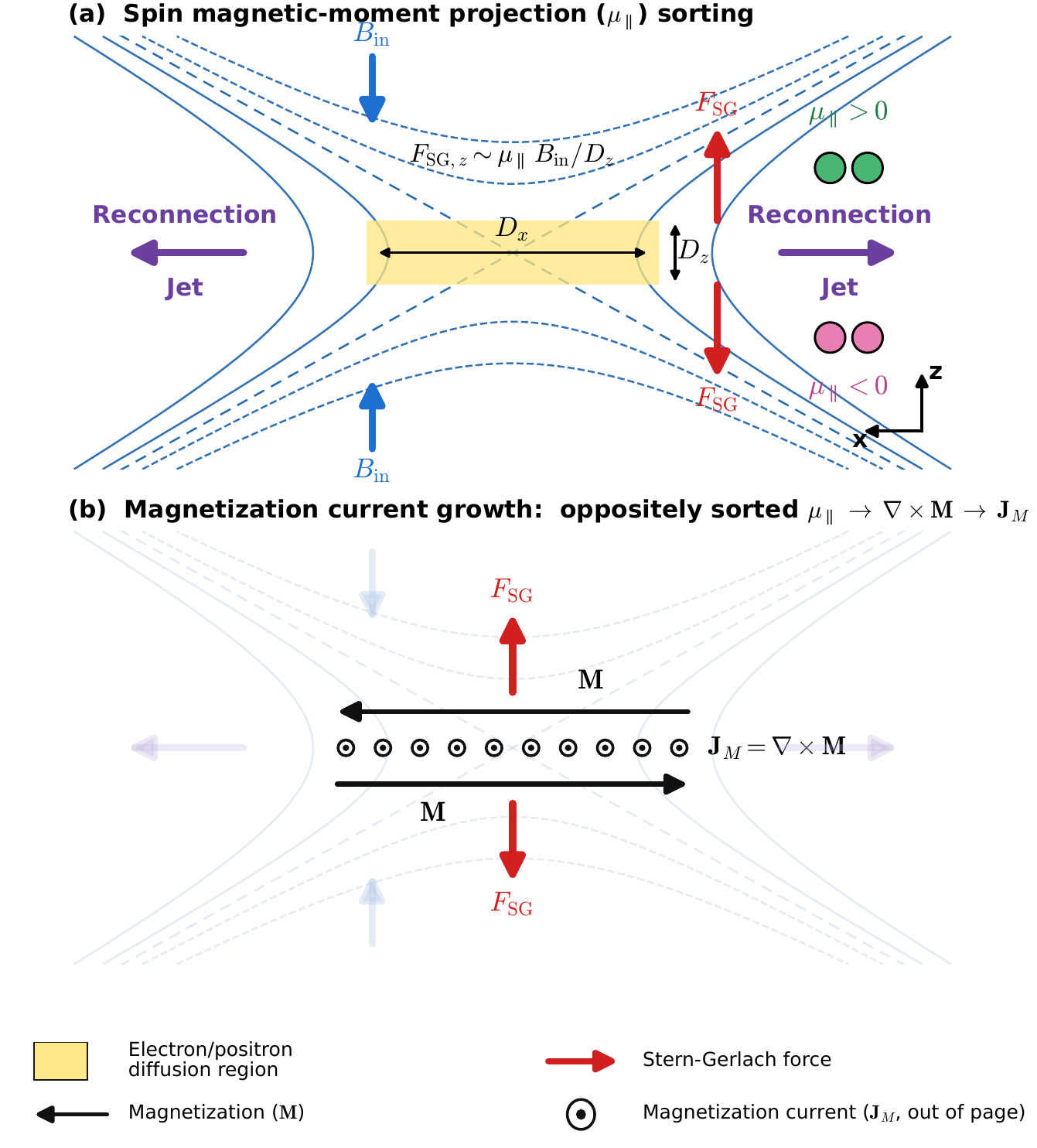}
\caption{\textbf{Schematic of SG sorting and magnetization-current development.} (a) Reconnection geometry and magnetic-moment sign convention, adapted from the companion Letter~\cite{NykyriPRLCompanion}.  The sheet-normal SG force separates particles according to the sign of $\mu_{\parallel}$.  (b) The oppositely sorted magnetic-moment branches produce a spatially varying magnetization $\mathbf M$ and hence an out-of-plane bound current $J_{M,y}=(\nabla\times\mathbf M)_y$.}
\label{fig:schematic}
\end{figure}

The production runs are deliberately focused on the seeded onset and early nonlinear response, where the matched cases can be compared
before their island-merger and saturation histories diverge: the normalized rate is measured over $3\le t/\tau_{\rm sp}\le7$, and the morphology is followed through $t=12\tau_{\rm sp}$. By the end of this interval, the sorting, pressure response, magnetization current, and flux-growth enhancement
are already established. The long classical control-run in Fig.~S9 of the Supplemental Material exhibits no measurable energy increase, confirming that the classical PIC implementation conserves energy over the simulated interval.

Collisionless reconnection rates are usually compared through the Alfv\'enic normalization $R_A=E_{\rm rec}/(B_{\rm up}V_{A,\rm up})$. Nonrelativistic electron--ion simulations, rate theory, and observations from magnetosphere multiscale (MMS) space mission commonly place the local normalized rate in the range $\sim0.1$--$0.2$~\cite{Shay2001,Liu2017,LiuCassak2022Rate,Genestreti2018,Pritchard2023}. Relativistic pair-plasma studies instead organize the upstream state primarily by the magnetization $\sigma$; fully kinetic simulations and first-principles theory likewise find fast rates of order $0.1$, with pressure depletion and open exhaust geometry playing central roles~\cite{Bessho2005,Liu2015,GoodbredLiu2022,Kagan2015}. Relativistic electron--proton PIC simulations report rates near $0.1$ over a broad range of ion magnetization, while alternative relativistic Alfv\'en-speed normalizations give values $0.14$--$0.25$~\cite{Werner2018,Melzani2014}. The parameters used here are complementary rather than
interchangeable with $\sigma$: $\gamma_{\rm tr}$ is the
characteristic transit parameter entering $\Xi$, whereas $\Xi$
compares the SG cross-sheet displacement with the relativistic
Larmor radius. In the relativistic sweep, the particle distribution
is explicitly loaded to realize the selected $\gamma_{\rm tr}$;
in the fixed-loading $\Xi$ scan, the Harris distribution is held fixed.

 Conventional spinless studies correspond to $\Xi=0$. Our fixed-$\gamma_{\rm tr}$ scan therefore asks whether adding spin transport changes flux growth relative to the same classical PIC state, while the fixed-$\chi$ relativistic family asks how that additional spin channel is suppressed as $\gamma_{\rm tr}$ increases. To compare with existing reconnection studies, we therefore test whether SG-induced branch anisotropy modifies reconnection through pressure-tensor, current, and nonideal-field channels, and we identify the limits of applicability of the semiclassical model as the field approaches the QED regime. The Supplemental Material provides implementation details, convergence and energy ledgers, the temporal ordering of the physical
mechanisms affecting reconnection, an Ohm's-law audit, and a strong-field applicability analysis. 
 
\section{SpinPIC2D model and numerical update}
\label{sec:model}

\subsection{Governing particle, spin, and field equations}

SpinPIC2D is a 2.5-dimensional pair-plasma particle-in-cell model built on the standard electromagnetic PIC architecture~\cite{Birdsall1985,Yee1966}. The simulation plane is $(x,z)$, where $x$ is the reconnection outflow direction and $z$ is the inflow/current-sheet normal; $\partial/\partial y=0$, but all three components of particle momentum, spin, current, and electromagnetic field are retained. Each macro-particle carries position $\mathbf x_i$, proper momentum $\mathbf p_i=\gamma_i m_e\mathbf v_i$, charge $q_i$, weight $w_i$, and a unit spin vector $\mathbf s_i$. Its equations are
\begin{align}
\frac{d\mathbf x_i}{dt}&=\frac{\mathbf p_i}{\gamma_i m_e},\\
\frac{d\mathbf p_i}{dt}&=q_i\left(\mathbf E+\mathbf v_i\times\mathbf B\right)+\mathbf F_{{\rm SG},i},
\label{eq:particle}\\
\frac{d\mathbf s_i}{dt}&=\bm\Omega_{{\rm BMT},i}\times\mathbf s_i,
\label{eq:bmt}
\end{align}
with
\begin{equation}
\mathbf F_{{\rm SG},i}=\nabla(\bm\mu_i\cdot\mathbf B),
\qquad
\bm\mu_i=g\frac{q_i}{2m_e}\mathbf S_i.
\label{eq:sgforce}
\end{equation}
The production branch uses the leading magnetic Bargmann--Michel--Telegdi rotation about the locally gathered field~\cite{BMT1959}. The full covariant BMT electric-field and anomalous-moment corrections are not included; this limitation is addressed in Sec.~\ref{sec:qed} and the Supplemental Material.

The particle free current and spin magnetization are deposited with the same area-weighting shape function $S$,
\begin{align}
\Jfree(\mathbf x)&=\sum_i q_i w_i\mathbf v_i S(\mathbf x-\mathbf x_i),\\
\Mfield(\mathbf x)&=\sum_i w_i\bm\mu_i S(\mathbf x-\mathbf x_i),
\end{align}
and the bound current is
\begin{equation}
\Jm=\nabla\times\Mfield.
\label{eq:jm}
\end{equation}
Maxwell's equations are advanced as
\begin{align}
\frac{\partial\mathbf B}{\partial t}&=-\nabla\times\mathbf E,\\
\frac{\partial\mathbf E}{\partial t}&=c^2\nabla\times\mathbf B-\frac{\Jfree+\Jm}{\epsilon_0}.
\label{eq:maxwell}
\end{align}
Thus $\Jm$ is not a post-processing diagnostic: in the full production branch it contributes directly to the electric-field update. The kinetic pressure tensor is likewise not imposed as a fluid closure; it emerges from the evolved particle distribution.

\subsection{Discrete update sequence and magnetization regularization}

The electromagnetic fields are stored on a Yee mesh~\cite{Yee1966}. At each time step the code (i) advances $\mathbf B$ by a half Faraday step, (ii) constructs the field view used by spin precession and the SG kick, (iii) applies a half-step BMT rotation, a full SG proper-momentum kick, the relativistic Boris Lorentz push~\cite{Boris1970}, the second half-step spin rotation, and the position/boundary update, (iv) deposits charge, free current, pressure moments, and component-staggered magnetization, (v) evaluates $\Jm=\nabla\times\Mfield$, (vi) advances $\mathbf E$ through Ampere's law using $\Jfree+\Jm$, and (vii) completes the second half Faraday step. This leapfrog ordering keeps the particle current at the half time level used by Ampere's law. The source-level staging, interpolation, compile flags, and energy-work ledgers are documented in the Supplemental Material.

Because Eq.~(\ref{eq:jm}) differentiates a particle-deposited moment, an unregularized curl amplifies grid-scale sampling noise. The accepted production branch applies a symmetric finite-width coarse-graining operator $\mathcal L_\ell$ to the deposited magnetization before taking the curl,
\begin{equation}
\Jm=\nabla\times(\mathcal L_\ell\Mfield),
\qquad \ell=1.10\de,
\label{eq:filter}
\end{equation}
with a fixed physical length rather than a fixed number of grid cells. Figure~\ref{fig:highk} and Figs.~S1--S3 show why this regularization is required and verify that the accepted branch preserves the sheet-scale response and energy closure.

\subsection{Microscopic spin coupling and SG ordering parameter}

The source code contains an input variable named \texttt{chi}.  To distinguish
that numerical control from its direct physical interpretation, we denote it
here by $\chi_{\rm sim}$.  The corresponding physical dimensionless quantum
coefficient is
\begin{equation}
\chi_{\rm phys}\equiv\frac{\hbar\omega_{pe}}{m_ec^2}.
\label{eq:chip}
\end{equation}
For a direct dimensional mapping one sets $\chi_{\rm sim}=\chi_{\rm phys}$.
More generally the code input can be written
$\chi_{\rm sim}=\kappa_\chi\chi_{\rm phys}$, where $\kappa_\chi$ is a
controlled similarity multiplier.  Within the fixed-loading reference scan, $\chi_{\rm sim}$ is varied
while the Harris particle distribution, grid, and seed are held fixed.
In the relativistic families, the initialized particle distribution is
changed by construction, and $\chi_{\rm sim}$ is either held fixed or
retuned to maintain $\Xi=1$. Consequently the numerical values of
$\chi_{\rm sim}$ are coupling coefficients used to probe the model but 
they are not, by themselves, direct estimates of
$\hbar\omega_{pe}/m_ec^2$ for a particular astrophysical environment.

In the adopted normalization $m_e=e=c=\omega_{pe}=k_B=1$, the magnetic-moment
coefficient is $\mu_{B,{\rm code}}=\chi_{\rm sim}/2$, and the SG term applied
to a particle carrying a unit spin vector $\mathbf s_i$ is
\begin{equation}
\left.\frac{d\mathbf p_i}{dt}\right|_{\rm SG}
=q_i\frac{g_e}{2}\frac{\chi_{\rm sim}}{2}
\,\nabla(\mathbf s_i\!\cdot\!\mathbf B).
\label{eq:sgchi}
\end{equation}
The same $\chi_{\rm sim}$ multiplies the deposited physical magnetization
before $\mathbf J_M=\nabla\times\mathbf M$ is evaluated.  Thus
$\chi_{\rm sim}$ controls the absolute spin-moment strength in both the
particle-force and Maxwell closures.  Below, ``fixed $\chi$'' is shorthand
for fixed $\chi_{\rm sim}$.

The macroscopic kinematic ordering parameter is distinct from this local
coefficient.  It compares the SG cross-sheet displacement accumulated during
a diffusion-region transit with the relativistic Larmor radius.  For inflow
field $\Bin$, current-sheet gradient scale $\Dz$, diffusion-region
half-length $\Dx$, transit speed $\btr c$, and characteristic Lorentz factor
$\avg{\gamma}$,
\begin{equation}
\XiZero = \frac{1}{8}\frac{\mu_B e}{m_e^2c^3}
\frac{\Bin^2 \Dx^2}{\avg{\gamma}^2\Dz}\,\btr^{-3},
\qquad
\XiEff=\Peff\XiZero.
\label{eq:xi0}
\end{equation}
Here $\Peff=\langle C\eta_\mu\rangle_n$ is the density-weighted effective branch
factor, combining coherent transit participation $C$ and the surviving
magnetic-moment projection $\eta_\mu=|\mu_\parallel|/\mu_B$. In code units,
\begin{equation}
\Xi=\frac{\Peff\chi_{\rm sim} B_{\rm in}^2D_x^2}
{16\gamma_{\rm tr}^2D_z\beta_{\rm tr}^3}.
\label{eq:xicode}
\end{equation}
Equation~(\ref{eq:xicode}) separates the two relativistic paths used below.
At fixed $\chi_{\rm sim}$, increasing $\gamma_{\rm tr}$ suppresses $\Xi$
through $(\gamma_{\rm tr}^2\beta_{\rm tr}^3)^{-1}$. Conversely, maintaining
$\Xi=1$ at fixed field and geometry requires
$\chi_{\rm sim}\propto\gamma_{\rm tr}^2\beta_{\rm tr}^3$; the target sequence
$\gamma_{\rm tr}=2,3,5$ uses $\chi_{\rm sim}=6.343$, $18.414$, and $57.410$,
respectively. This path preserves the transit-scale displacement ratio while
increasing the absolute magnetic-moment and magnetization-current coupling.
Fixed $\Xi$ therefore need not imply a common nonlinear state. 
Astrophysical regime estimates are organized by $\Xi_{\rm eff}$, with the adopted physical
$\chi_{\rm phys}$ and participation factor stated separately.

\section{Relativistic sorting, magnetization current, and magnetic response}
\label{sec:fig1physics}

Figure~\ref{fig:schematic} summarizes the microscopic mechanism.  In panel (a), the dominant Harris-sheet gradient produces $F_{{\rm SG},z}\simeq\mupar\Bin/\Dz$, so particles sort by the sign of $\mupar$ rather than by charge.  Panel (b) shows the corresponding electromagnetic feedback: opposite $\mupar$ branches create a spatially varying magnetization $\mathbf M$, whose curl produces the out-of-plane bound current $J_{M,y}=(\nabla\times\mathbf M)_y$ that enters Ampere's law.

\begin{figure*}[t]
\centering
\safeincludegraphics[width=0.99\textwidth]{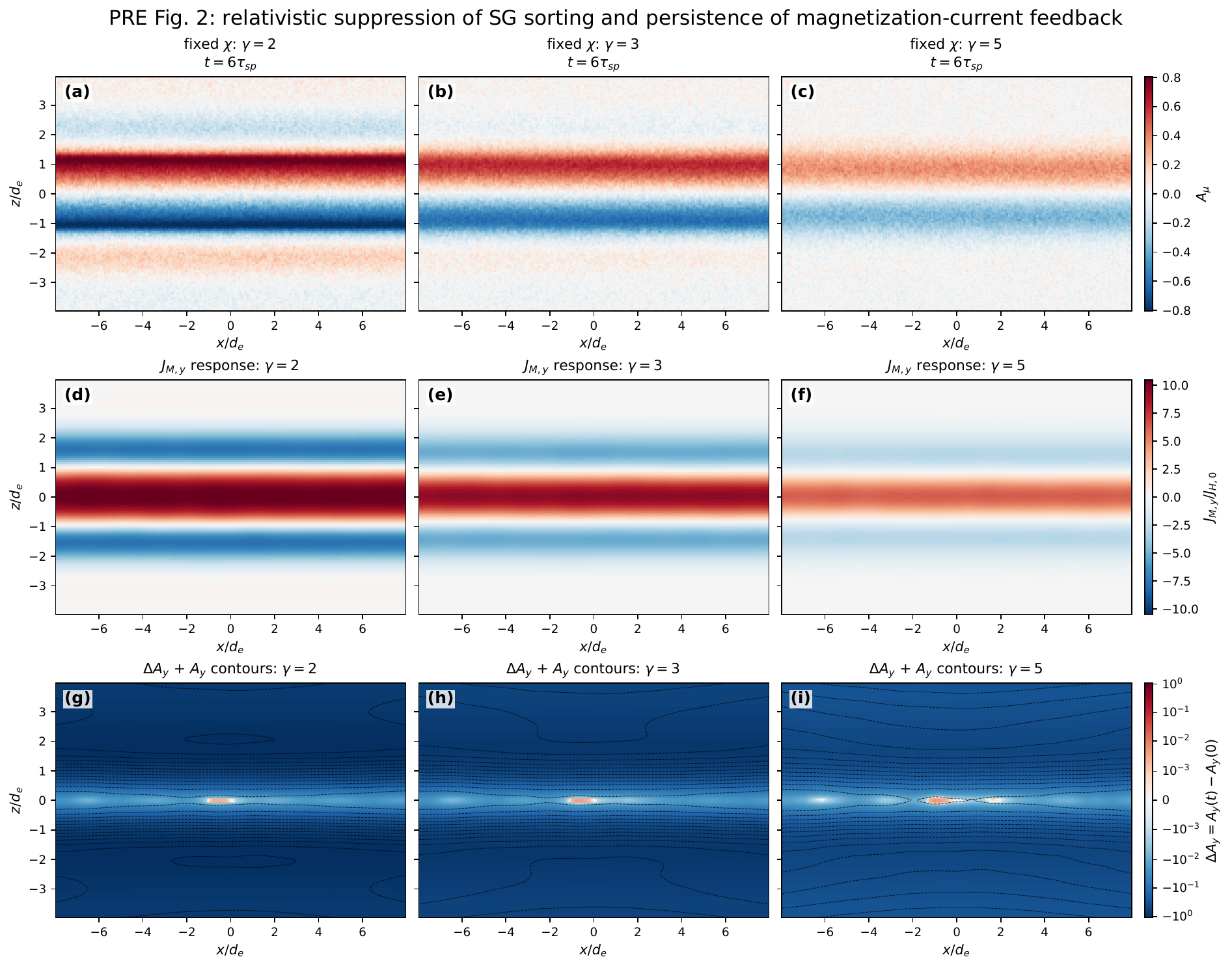}
\caption{\textbf{SG sorting, magnetization current, and magnetic response across the relativistic fixed-$\chi$ branch.} (a)--(c) Branch asymmetry $A_\mu$ at the same normalized time for fixed-$\chi$ simulations with target transit Lorentz factors $\gamma_{\rm tr}=2$, 3, and 5. (d)--(f) Corresponding $J_{M,y}/J_{H,0}$ maps. (g)--(i) $\Delta A_y=A_y(t)-A_y(0)$ with black contours of total reconstructed $A_y$.}
\label{fig:relativistic_maps}
\end{figure*}

Figure~\ref{fig:relativistic_maps} shows the three response channels at the common time $t=6\tausp$ for fixed-$\chi$ runs with target $\gamma_{\rm tr}=2,3,5$. Panels (a)--(c) show the branch-density asymmetry $A_\mu$ on a common color scale. Panels (d)--(f) show the sheet-normalized magnetization current $J_{M,y}/J_{H,0}$, and panels (g)--(i) show the magnetic response $\Delta A_y$ with total-$A_y$ contours. The branch asymmetry decreases most clearly with increasing $\gamma_{\rm tr}$, whereas structured $J_{M,y}$ and $\Delta A_y$ remain visible through $\gamma_{\rm tr}=5$. Here $J_{H,0}=B_0/\lambda$ is the Harris current scale in code units. The apparent difference between Figs.~\ref{fig:relativistic_maps} and~\ref{fig:relativistic}(c) results from the current normalization: Fig.~\ref{fig:relativistic_maps} uses $J_{M,y}/J_{H,0}$, whereas Fig.~\ref{fig:relativistic}(c) shows the raw p99 amplitude. The exact definitions of $A_\mu$, p99, $J_{H,0}$, and $\Delta A_y$ are collected in Appendix~\ref{app:diagnostics}.

\section{Initialization and Harris equilibrium}
\label{sec:initialization}

\begin{figure*}[t]
\centering
\safeincludegraphics[width=0.92\textwidth]{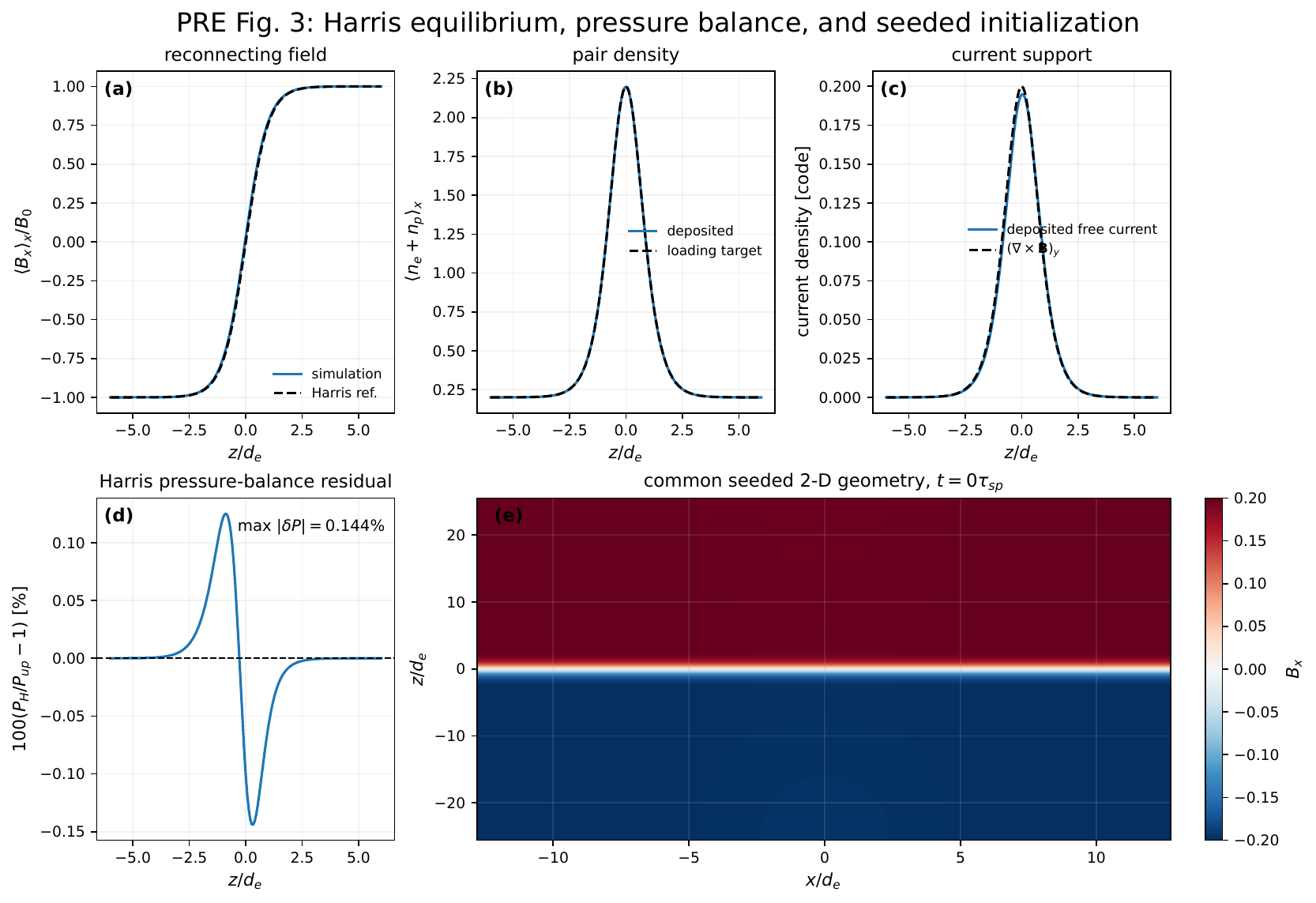}
\caption{\textbf{Harris equilibrium, pressure balance, and seeded initialization.} (a) $\avg{B_x}_x/B_0$ compared with the analytic Harris profile. (b) Deposited $\avg{n_e+n_p}_x$ compared with the loading target. (c) Deposited free current and the backward-Yee $(\nabla\times\mathbf B)_y$. (d) Residual $100(P_H/P_{\rm up}-1)$, where $P_H=n_{\rm pair}T+B_H^2/2$ uses the code-prescribed thermal pressure and the unperturbed Harris field. (e) Common weakly seeded two-dimensional $B_x$ geometry at $t=0$. The $0.144\%$ residual quantifies the underlying one-dimensional initialization audit; the finite seed is shown separately in panel (e).}
\label{fig:harris}
\end{figure*}

The production runs use a single pair-plasma Harris sheet~\cite{Harris1962} with a common weak flux perturbation that establishes the initial X-line saddle. Figure~\ref{fig:harris}(a) shows that the $x$-averaged reconnecting field follows $B_x/B_0=\tanh(z/\lambda)$, while panel (b) verifies the deposited pair-density profile and panel (c) compares the deposited free current with the backward-Yee curl of the initialized magnetic field.

Panel (d) evaluates the underlying one-dimensional Harris balance using
the same thermodynamic prescription as the initialization audit,

\begin{equation}
\begin{aligned}
P_H(z) &= n_{\rm pair}(z)\,T + \frac{B_H^2(z)}{2}, \\
T &= \frac{\beta}{2}, \\
B_H(z) &= B_0 \tanh\!\left(\frac{z}{\lambda}\right).
\end{aligned}
\end{equation}

where temperature is expressed in energy units because $k_B=1$. The maximum residual is $0.144\%$. Panel (e) separately shows the common finite two-dimensional perturbation that establishes the central X-line saddle. These checks show that the systematic $\Xi$ dependence is not caused by different macroscopic Harris loadings or seed geometries.

\section{Magnetization-current regularization}
\label{sec:regularization}

\begin{figure*}[t]
\centering
\safeincludegraphics[width=0.95\textwidth]{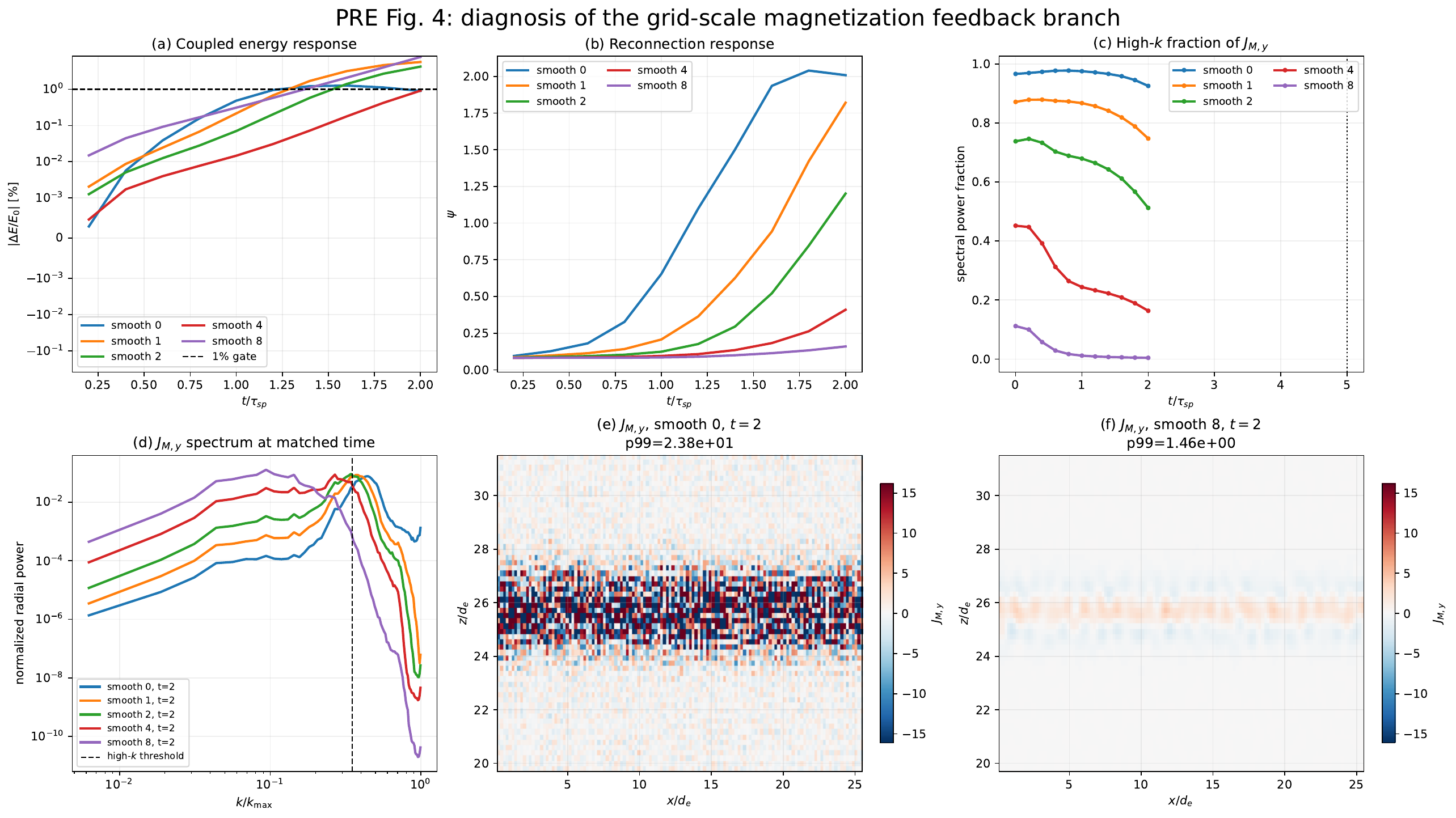}
\caption{\textbf{Diagnosis of the grid-scale magnetization-feedback branch for the $\Xi=1$-case.} (a) Coupled energy response for increasing smoothing strength, with the one-percent gate. (b) Reconnected-flux response. (c) High-$k$ fraction of $J_{M,y}$ spectral power. (d) Radially averaged $J_{M,y}$ power spectra at matched time. (e) Unsmoothed and (f) accepted strongly smoothed current maps at $t=2\tausp$. The accepted physical-width filter removes the high-$k$ branch while retaining the sheet-scale magnetization current.}
\label{fig:highk}
\end{figure*}

Figure~\ref{fig:highk} isolates the numerical branch created when the curl in Eq.~(\ref{eq:jm}) acts directly on noisy deposited magnetization for the $\Xi=1$ case. With no or weak smoothing, the coupled energy error and apparent reconnection response rise rapidly [panels (a) and (b)], most of the $J_{M,y}$ power occupies the high-$k$ portion of the spectrum [panels (c) and (d)], and the current map exhibits alternating grid-scale structure [panel (e)]. Increasing the fixed physical filter length removes that branch and leaves a smooth sheet-scale current [panel (f)]. The production filter is selected from the combined spectral, energy-drift, and physical-scale criteria.

\section{Production validation and robustness}
\label{sec:validation}
\begin{figure*}[t]
\centering
\safeincludegraphics[width=0.92\textwidth]{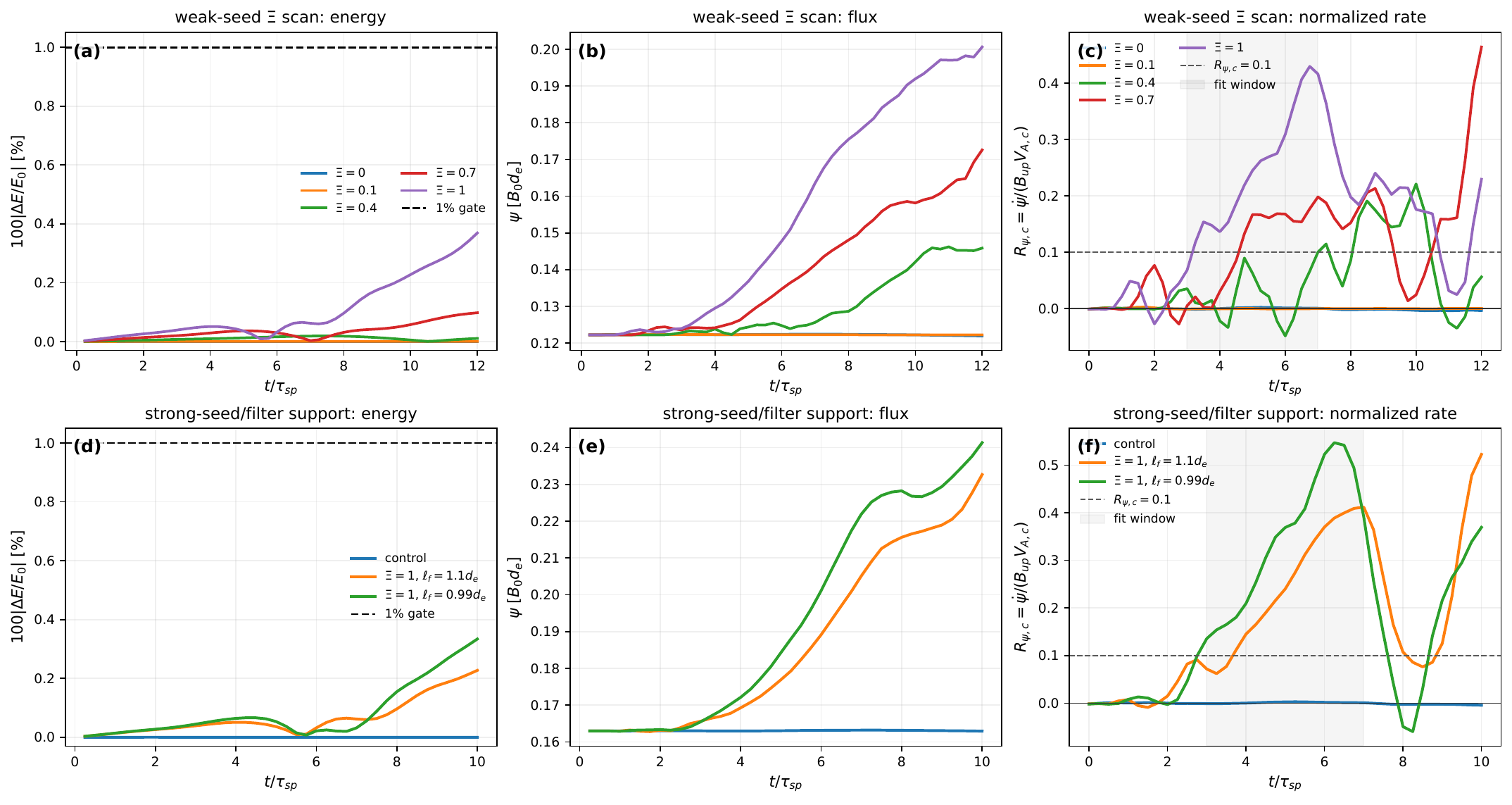}
\caption{\textbf{Production validation, energy gate, and normalized flux-growth robustness.} (a)--(c) Common weak-seed $\Xi=0,0.1,0.4,0.7,1$ production family: total-energy drift, global peak-to-peak flux $\psi(t)$, and the cold-rest-mass-normalized global flux-growth proxy $R_{\psi,c}=\dot\psi/[B_{\rm up}V_{A,c}]$. (d)--(f) The same diagnostics for the strong-seed control and the two neighboring accepted magnetization-filter lengths. The dashed lines in (a) and (d) mark the 1\% energy gate; the shaded intervals in (c) and (f) identify the common $3\le t/\tau_{sp}\le7$ fit window. The upstream field and total pair density are measured separately in each snapshot from $|x|\le6d_e$ and $4d_e\le|z|\le12d_e$, with $V_{A,c}/c=[\sigma_c/(1+\sigma_c)]^{1/2}$ and $\sigma_c=B_{\rm up}^2/(\mu_0 n_{\rm pair,up}m_ec^2)$. Once multiple X and O points develop, the plotted quantity is not a single-X-line reconnection rate.}
\label{fig:validation}
\end{figure*}

Figures~\ref{fig:validation} and~\ref{fig:xiscan} analyze the same
common-$\gamma_{\rm tr}=2$ weak-seed $\Xi$ production scan reported in
the companion Letter~\cite{NykyriPRLCompanion}. Here we provide the
energy validation, normalization details, filter-sensitivity tests, and
additional diagnostics underlying that result.

The fixed-$\gamma_{\rm tr}=2$ reference scan uses the common Harris
thermal loading, with $\gamma_{\rm tr}$ serving as the characteristic
transit parameter entering the definition of $\Xi$. The measured mean
initial particle Lorentz factor in this family is approximately
$\langle\gamma\rangle_0=1.015$.

Figure~\ref{fig:validation} applies the same acceptance criterion and
flux-growth normalization to all runs in the $\Xi$ run-family. Panels (a)--(c) show the
weak (0.075)-seed $\Xi=0,0.1,0.4,0.7,1$ scan. Every run remains below the one-percent
maximum total-energy-drift gate through $12\tausp$ [panel (a)], while the flux
histories separate systematically [panel (b)]. Over the common
$3\le t/\tausp\le7$ fit interval, the cold rest-mass-normalized global
flux-growth approximation $R_{\psi,c}=\dot\psi/(B_{\rm up}V_{A,c})$ remains close to
the control-run for $\Xi\le0.1$, is $0.016$ at $\Xi=0.4$, and reaches $0.12$ and
$0.25$ at $\Xi=0.7$ and 1, respectively [panel (c)]. Because
$B_{\rm up}V_{A,c}$ varies by less than one percent across the scan, this
ordering originates in $\dot\psi$ rather than in a changing normalization.
The magnitude can be compared with conventional reconnection rates,
but the diagnostic represents a local X-line rate only while a single
X--O pair dominates.

Panels (d)--(f) repeat the test for a stronger (0.1)-seed and two selected filter lengths. 
The control remains quiescent, whereas both
$\Xi=1$ runs exhibit rapid flux growth. Their fitted amplitudes differ at the
order-unity level, indicating finite sensitivity of the nonlinear magnitude
to the magnetization coarse-graining length. The timestep, grid,
particle-number, and work-channel tests in the Supplemental Material extend
this comparison beyond a single seed and discretization. After multiple extrema form, the plotted quantity is a global,
normalized approximation to the topology-growth rate rather than a
unique local X-line rate.

\section{Validated $\Xi$ scaling of sorting and reconnection feedback}
\label{sec:xiscan}

%The reference $\Xi$ scan reported in the companion
%Letter~\cite{NykyriPRLCompanion} keeps the Harris particle loading
%fixed and uses $\gamma_{\rm tr}=2$ as the transit parameter in the
%definition of $\Xi$. The relativistic sweep instead changes the
%nitialized particle distribution directly, giving measured initial
%values $\langle\gamma\rangle_0\simeq2.01$, 3.01, and 5.01. Its
%owest-energy case is therefore not a repeat of the Letter's A4 run,
%although both use $\Xi=1$ and $\chi_{\rm sim}=6.343$. The two cases
%have different particle distributions and relativistic inertia.

\begin{figure*}[t]
\centering
\safeincludegraphics[width=0.94\textwidth]{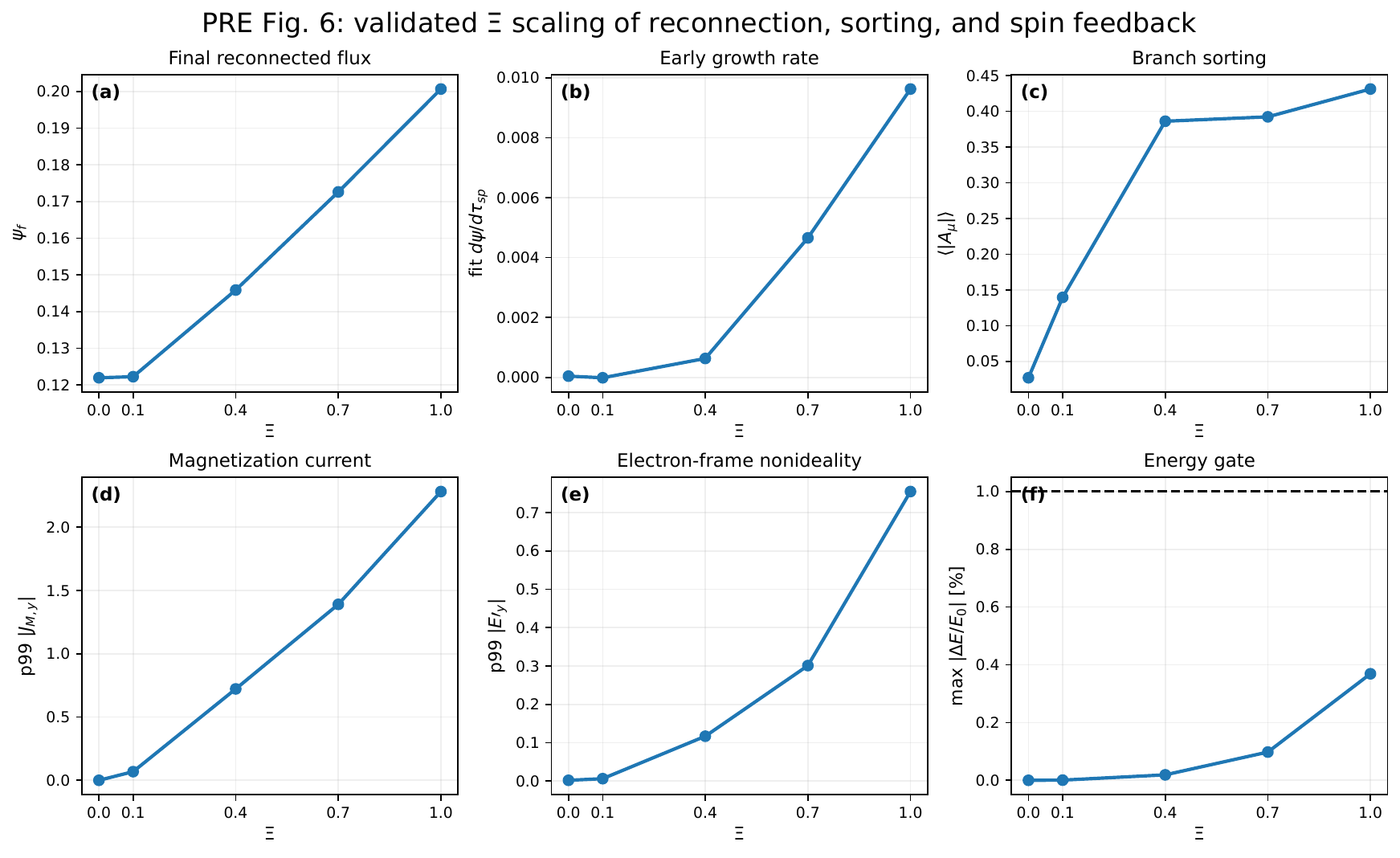}
\caption{\textbf{Validated $\Xi$ scaling of reconnection, sorting, and spin feedback.} (a) Final flux $\psi_f$. (b) Early fitted $d\psi/d\tausp$. (c) Mean absolute branch asymmetry $\avg{|A_\mu|}$. (d) p99 $|J_{M,y}|$. (e) p99 electron-frame nonideal field $|E'_y|$. (f) Maximum total-energy drift $100|\Delta E/E_0|$. All quantities are shown for the fixed-$\gamma_{\rm tr}=2$ scan. Spatial metrics use the common X-line box $|x|\le1.5\de$, $|z|\le1.0\de$ at $t\simeq7\tausp$; p99 denotes the 99th percentile of the absolute finite cell values. The branch asymmetry, magnetization-current amplitude, electron-frame nonideality, early flux growth, and final flux follow the same ordering with $\Xi$.}
\label{fig:xiscan}
\end{figure*}

Figure~\ref{fig:xiscan} summarizes six diagnostics from the common $\gamma_{\rm tr}=2$ production scan. The flux, branch-asymmetry, magnetization-current, and electron-frame nonideality measures remain near the control for $\Xi\le0.1$, begin to increase at $\Xi=0.4$, and rise strongly at $\Xi=0.7$ and 1, while the energy drift remains below 1\%. Their common ordering supports a dynamical association between SG sorting, current restructuring, and enhanced flux growth. The plotted slope is used as a matched-run flux-growth diagnostic rather than equated numerically with the conventional Alfv\'en-normalized rate; Appendix~\ref{app:diagnostics} gives the conversion and the conditions under which it approaches a local X-line rate.

\section{Relativistic population dependence and limits of single-parameter similarity}
\label{sec:relativistic}

The reference $\Xi$ scan reported in the companion
Letter~\cite{NykyriPRLCompanion} keeps the Harris particle loading
fixed and uses $\gamma_{\rm tr}=2$ as the transit parameter entering
the definition of $\Xi$. The relativistic sweep addresses a different
question by changing the initialized particle distribution directly,
with measured initial values
$\langle\gamma\rangle_0\simeq2.01$, 3.01, and 5.01. Its
lowest-energy case is therefore not a repeat of the Letter's A4 run,
although both use $\Xi=1$ and $\chi_{\rm sim}=6.343$. The two cases
have different particle distributions and relativistic inertia.

\begin{figure*}[t]
\centering
\safeincludegraphics[width=0.96\textwidth]{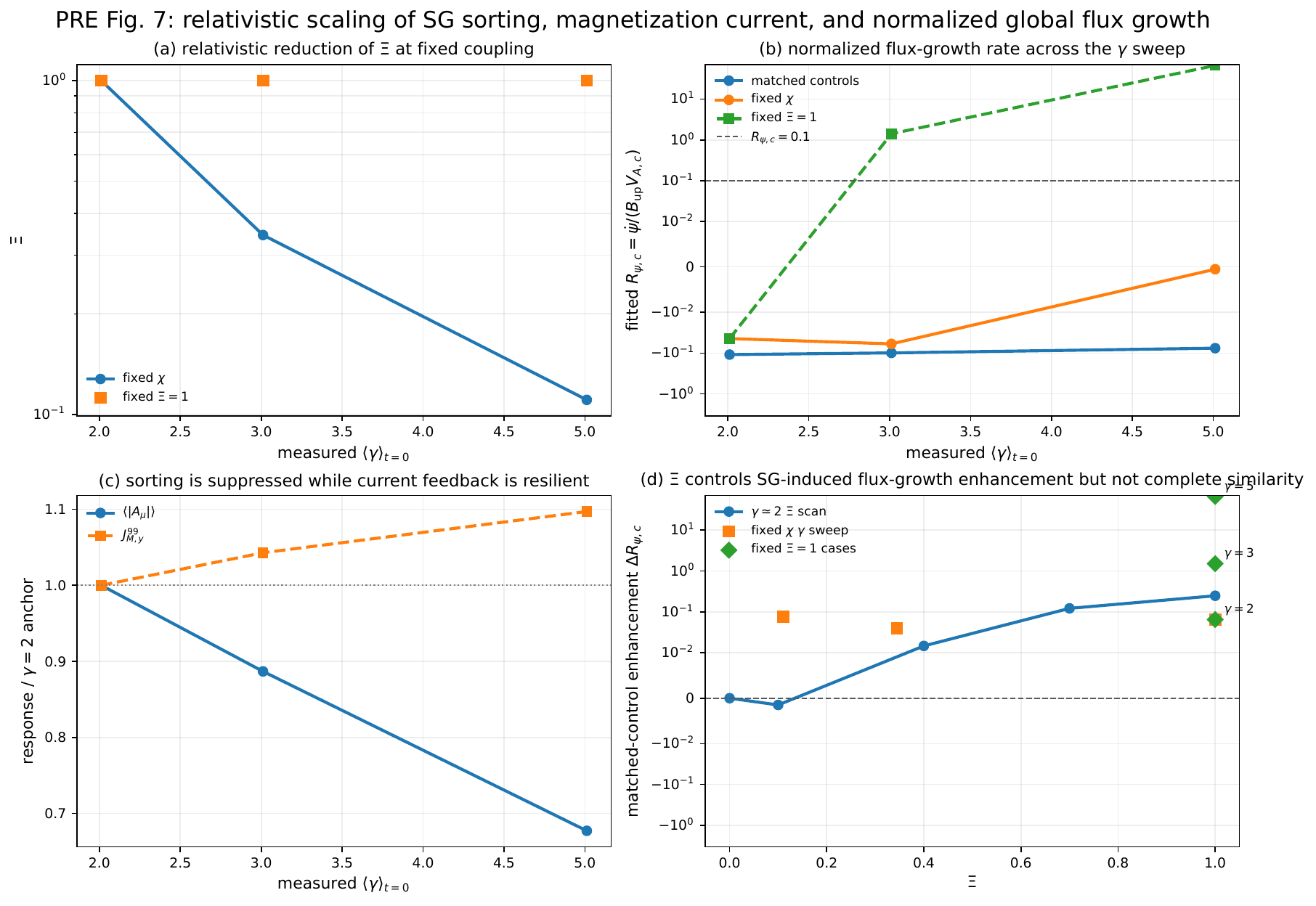}
\caption{\textbf{Relativistic scaling of SG sorting, magnetization current, and normalized global flux growth.} The horizontal coordinate is the measured $\avg{\gamma}_{t=0}$, the realized value of the target $\gamma_{\rm tr}=2,3,5$ sequence. (a) $\Xi$ along the fixed-$\chi$ branch and the separately tuned fixed-$\Xi=1$ cases. (b) Fitted cold-rest-mass-normalized global flux-growth proxy $R_{\psi,c}=\dot\psi/(B_{\rm up}V_{A,c})$ over $3\le t/\tau_{sp}\le7$ for matched controls, fixed-$\chi_{\rm sim}$ runs, and fixed-$\Xi=1$ runs. Negative control values reflect relaxation of the imposed peak-to-peak seed. (c) Fixed-$\chi$ sorting and raw p99 $J_{M,y}$, each normalized to the $\gamma_{\rm tr}=2$ anchor. (d) Matched-control enhancement $\Delta R_{\psi,c}=R_{\psi,c}^{\rm full}-R_{\psi,c}^{\rm ctrl}$ for the reference $\gamma_{\rm tr}\simeq2$ $\Xi$ scan, the relativistic fixed-$\chi$ sweep, and the fixed-$\Xi=1$ cases. The $\gamma_{\rm tr}=2$ anchor in this figure is the explicitly loaded relativistic run \texttt{REL\_G2\_FULL\_fixedChi}, with measured $\langle\gamma\rangle_0\simeq2.01$. It is distinct from the A4 $\Xi=1$ endpoint of the reference scan, which uses the common Harris thermal loading. The refined $\gamma_{\rm tr}=5,\Xi=1$ run uses $\Delta t=6\times10^{-4}$ and satisfies $\max|\Delta E/E_0|=0.459\%$. Its large value is a global multi-X-line topology-growth approximation, not the local reconnection rate of a single X line.}
\label{fig:relativistic}
\end{figure*}

Figure~\ref{fig:relativistic} separates the kinematic SG ordering from
the absolute electromagnetic feedback strength. Along the
fixed-$\chi_{\rm sim}$ path, panel (a) recovers the analytic reduction
$\Xi\propto(\gamma_{\rm tr}^2\beta_{\rm tr}^3)^{-1}$, and panel (c)
shows that the direct sorting amplitude falls to approximately $0.89$
and $0.68$ of the $\gamma_{\rm tr}=2$ value at
$\gamma_{\rm tr}=3$ and 5. In contrast, the raw p99
magnetization-current amplitude reaches approximately $1.04$ and
$1.10$ of the $\gamma_{\rm tr}=2$ anchor. Because
$\mathbf J_M=\nabla\times\mathbf M$ depends on spatial gradients as
well as the scalar sorting amplitude, weaker visible branch separation
does not imply proportionally weaker current feedback.

Panel (b) shows the fitted cold-normalized global flux-growth approximation for the
matched controls, fixed-$\chi_{\rm sim}$ runs, and fixed-$\Xi=1$ runs. The
controls have negative slopes during $3\le t/\tausp\le7$ because the imposed
peak-to-peak seed relaxes; these values do not represent reverse reconnection.
Subtracting each matched control gives positive fixed-$\chi_{\rm sim}$
enhancements $\Delta R_{\psi,c}\simeq0.065$, $0.039$, and $0.075$ for
$\gamma_{\rm tr}=2,3,5$, respectively. The matched-control enhancement
therefore remains positive along the relativistic sequence even as direct
sorting decreases.

Along the separately tuned fixed-$\Xi=1$ path,
$\chi_{\rm sim}$ increases from $6.343$ to $18.414$ to $57.410$, and the raw
p99 magnetization-current amplitude increases from approximately $1.89$ to
$7.76$ to $45.48$. The corresponding global flux-growth proxies are
approximately $-0.044$, $1.41$, and $66.35$. The value $66.35$ is not a local
Alfv\'en-normalized inflow rate because the sheet already contains multiple X
and O points; it measures growth and motion of the global flux extrema. Its
normalization denominator $B_{\rm up}V_{A,c}\simeq0.106$ is comparable to the
$\simeq0.08$ values of the other runs. Hence $\Xi$ orders access to the
SG-modified regime but does not alone specify the nonlinear state.

\section{Strong-coupling transition at $\gamma_{\rm tr}=5,\Xi=1$}
\label{sec:g5}

\begin{figure*}[t]
\centering
\safeincludegraphics[width=0.88\textwidth]{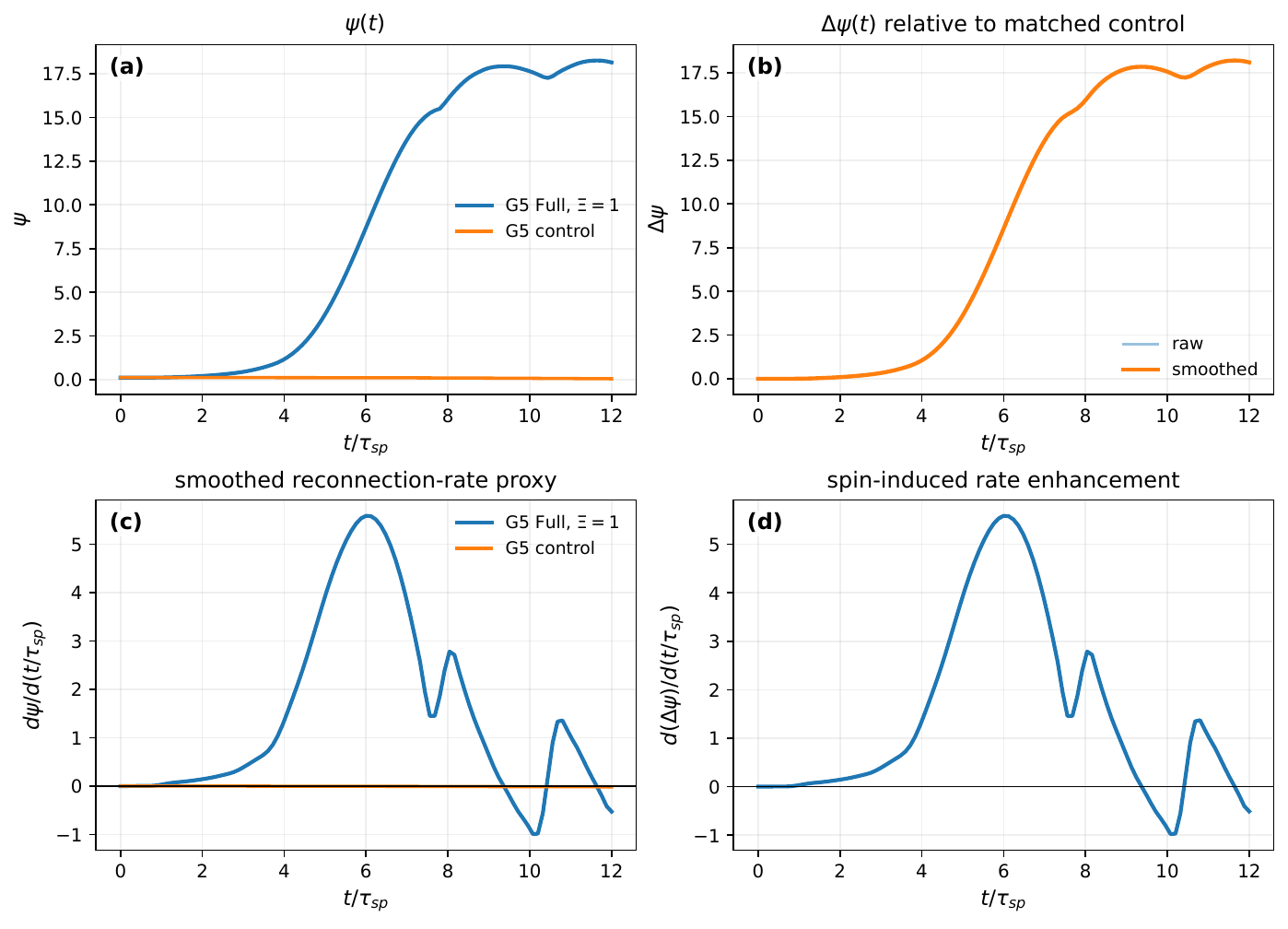}\\[1ex]
\safeincludegraphics[width=0.96\textwidth]{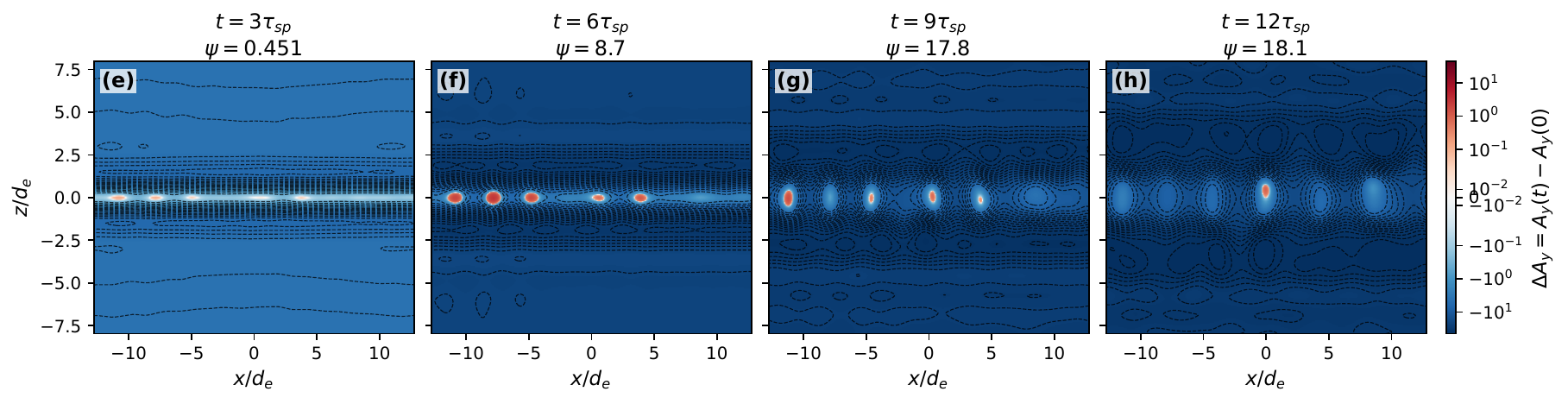}
\caption{\textbf{Transition to a nonlinear multi-X-line state in the $\gamma_{\rm tr}=5,\Xi=1$ regime.} (a) $\psi(t)$ for the full and matched-control runs. (b) $\Delta\psi(t)$ relative to the matched control. (c) Smoothed global $d\psi/d(t/\tausp)$ and (d) the spin-induced enhancement $d(\Delta\psi)/d(t/\tausp)$. (e)--(h) $\Delta A_y$ at $t/\tausp=3,6,9,12$, respectively, with black contours of total reconstructed $A_y$. The rapid global flux-growth interval coincides with breakup of the quasi-one-dimensional sheet into a localized multi-X-line topology.}
\label{fig:g5transition}
\end{figure*}

Figure~\ref{fig:g5transition} resolves the temporal evolution of the fixed-$\Xi=1$, $\gamma_{\rm tr}=5$ case. The full run remains close to the matched control through the early stage, then $\psi$ and $\Delta\psi$ rise rapidly between approximately $4$ and $8\tausp$ [panels (a) and (b)]. The smoothed global rate proxy peaks near $6\tausp$ [panels (c) and (d)]. At $3\tausp$ the topology remains close to a quasi-one-dimensional sheet [panel (e)]; by $6\tausp$ multiple localized X--O structures appear [panel (f)], by $9\tausp$ the global flux approaches saturation [panel (g)], and by $12\tausp$ the islands undergo late reorganization [panel (h)]. The rapid flux-growth interval coincides with the breakup of the quasi-one-dimensional sheet into a multi-X-line topology. After multiple extrema form, the rate diagnostic is a global peak-to-peak flux-growth approximation rather than the normalized reconnection electric field of one selected X-line.

\section{Discussion}

\subsection{How the SG anisotropy modifies reconnection}
\label{sec:dae}

The SG force is present in the unreconnected Harris sheet because $\partial_zB_x\ne0$. It first produces a moment-tagged north--south redistribution [Figs.~\ref{fig:schematic} and~\ref{fig:relativistic_maps}], which changes branch-resolved distributions and pressure moments. The kinetic pressure tensor is
\begin{equation}
\mathbf P_s=m_s\int (\mathbf v-\mathbf u_s)(\mathbf v-\mathbf u_s)f_s\,d^3v,
\label{eq:pressure}
\end{equation}
so spatially separated branches with different drift correlations can change both diagonal pressure and off-diagonal stress. Simultaneously, the sorted magnetic moments generate $\Mfield$ and $\Jm$, which enter the current system and Maxwell update. The common-$\Xi$ ordering in Fig.~\ref{fig:xiscan}, the onset chronology in Fig.~S4, and the generalized-Ohm's law maps in the companion
Letter~\cite{NykyriPRLCompanion} indicate that sorting and pressure/current restructuring precede or accompany the accelerated flux growth rather than appearing only after reconnection has saturated. This sequence is consistent with established rate theories in which X-line pressure support, current-carrier energetics, and exhaust opening control fast reconnection~\cite{Liu2017,LiuCassak2022Rate,GoodbredLiu2022}.

For electrons the diagnosed $y$-component of the momentum balance is
\begin{equation}
E'_y=\mathcal P_e+\mathcal I_e+\mathcal S_e+R_e,
\label{eq:ohmcompact}
\end{equation}
where $E'_y=E_y+(\mathbf u_e\times\mathbf B)_y$, $\mathcal P_e=(\nabla\cdot\mathbf P_e)_y/(q_en_e)$, $\mathcal I_e=(m_e/q_e)D_tu_{e,y}$, and $\mathcal S_e=-F_{{\rm SG},e,y}/(q_en_e)$. Because $\partial/\partial y=0$ in the 2.5-D geometry, $F_{{\rm SG},y}=0$ and the SG effect is indirect: it modifies $\mathbf P_e$, particle inertia, $\Mfield$, and $\Jm$ rather than supplying a direct $y$-directed electric-force term. The remaining structured, residual,  $R_e$,  is retained as a closure diagnostic; its possible current-dependent or wave-mediated interpretation is deferred to future work. These results identify a dynamical pathway through the pressure tensor and magnetization current. A linear spin-tearing analysis is beyond the scope of the present work.

\subsection{Relation to conventional pair- and electron--ion reconnection rates}
\label{sec:ratecontext}

The SG-induced enhancement reported here should be interpreted as an additional control on top of the familiar collisionless reconnection baseline, not as a replacement for the upstream magnetization $\sigma$. In conventional relativistic pair PIC, increasing $\sigma$ raises the relativistic Alfv\'en speed and can increase the dimensional inflow speed, while the Alfv\'en-normalized rate remains of order $0.1$ and the diffusion-region aspect ratio remains near $0.1$~\cite{Liu2015,Kagan2015}. Goodbred and Liu~\cite{GoodbredLiu2022} showed that, at large $\sigma$, the energy required to carry the X-line current depletes the thermal pressure, allowing the upstream field to collapse into an open exhaust and sustain fast reconnection. In nonrelativistic electron--ion plasmas, Hall-mediated diversion of electromagnetic energy produces an analogous pressure depletion and open geometry~\cite{Liu2017,LiuCassak2022Rate}. MMS measurements provide an observational benchmark: one well-resolved diffusion-region event yielded $R_A=0.18\pm0.035$~\cite{Genestreti2018}, and a multi-event magnetopause study found a typical range $0.1$--$0.2$ with mean value about $0.15$~\cite{Pritchard2023}.

Our relativistic family follows a different path through parameter
space. In the fixed-loading $\Xi$ scan of the companion
Letter~\cite{NykyriPRLCompanion}, $\gamma_{\rm tr}=2$ is the transit
parameter entering $\Xi$, while the Harris particle distribution is
held fixed. In the relativistic sweep, by contrast, the particle
distribution is explicitly loaded to measured initial values
$\langle\gamma\rangle_0\simeq2.01$, 3.01, and 5.01. Thus the
$\gamma_{\rm tr}=2$, $\Xi=1$ relativistic-sweep case is not a repeat
of the Letter's A4 run, despite having the same $\Xi$ and
$\chi_{\rm sim}$.

 At fixed microscopic coupling $\chi_{\rm sim}$, increasing $\gamma_{\rm tr}$ reduces $\Xi\propto(\gamma_{\rm tr}^2\beta_{\rm tr}^3)^{-1}$ and suppresses visible branch sorting even though conventional relativistic reconnection can remain fast as $\sigma$ increases. The parameter $\sigma$ controls the classical Alfv\'enic energy reservoir and outflow, whereas $\Xi$ controls the added SG displacement and its pressure/current feedback. The fixed-$\gamma_{\rm tr}=2$ scan isolates this added channel by holding the classical PIC state fixed while varying $\Xi$. The separately tuned fixed-$\Xi=1$ family shows that preserving the kinematic SG ratio does not preserve the nonlinear state because the required microscopic coupling and resulting magnetization current change with $\gamma_{\rm tr}$.

Relativistic electron--proton PIC provides a useful composition comparison. Werner et al.~\cite{Werner2018} found $R_A\simeq0.1$ from semirelativistic to fully relativistic ion magnetizations, whereas Melzani et al.~\cite{Melzani2014} obtained $0.14$--$0.25$ when normalizing by the projected relativistic inflow Alfv\'en speed. Normalized rates depend on the upstream and local normalization, and comparisons require consistent definitions of $B_{\rm up}$, $V_A$, and the X--O flux difference. In the controlled $\Xi$ scan of Fig.~\ref{fig:validation}, the cold-normalized global flux-growth approximation crosses the conventional $R_A\sim0.1$ scale between $\Xi=0.4$ and $0.7$ while the upstream normalization remains nearly unchanged. This comparison approaches a local rate only during the single-X--O stage; the strong-coupling multi-island values in Fig.~\ref{fig:relativistic} are global topology-growth estimates, as defined in Appendix~\ref{app:diagnostics}.

The present simulations use a symmetric, antiparallel Harris sheet with equal upstream pair densities and no guide field. In this
geometrical sense they are closer to idealized magnetotail reconnection, where the two inflow regions can be approximately symmetric, than to
dayside magnetopause reconnection. At the Earth's magnetopause (and other magnetized planets), the upstream densities and reconnecting magnetic fields generally differ, and the rate must be normalized using the corresponding asymmetric outflow or
Alfv\'en speed. The normalized values obtained here therefore should not be transferred directly to dayside magnetopause conditions.

A finite guide field would introduce an additional dependence through
the BMT spin precession, the magnetic-moment projection
$\mu_\parallel$, SG sorting, and the resulting magnetization current.
In the present 2.5-D geometry, however, adding a guide field would not
by itself produce $F_{{\rm SG},y}$ because
$\partial/\partial y=0$. A fully three-dimensional model would permit
out-of-plane gradients and hence a finite $F_{{\rm SG},y}$, while also
allowing oblique, drift-kink, and turbulence-mediated dynamics that are
excluded from the present simulations. Asymmetric inflow conditions,
finite-guide-field scans, and a three-dimensional extension of
SpinPIC2D are left for future work.
\subsection{Radiative spin selection and the QED boundary}
\label{sec:qed}

The semiclassical SG+BMT model should not be extrapolated without qualification to 
\[B_Q
= \frac{m_e^2 c^2}{e\hbar}
\simeq 4.414\times10^{9}\ {\rm T}
= 4.414\times10^{13}\ {\rm G},
\]
where Landau quantization and strong-field radiative processes become important~\cite{HardingLai2006}. The companion Letter \cite{NykyriPRLCompanion} discusses the astrophysical survey of  the $\Xi$-parameter, and Supplemental Material places this  survey in $(B/B_Q,\XiEff)$ space and compares the Sokolov--Ternov (ST) spin-flip time with the current-sheet transit time. With the adopted regime estimates, $\tau_{\rm ST}/T_r\simeq63$ in the magnetar magnetosphere, so radiative polarization does not complete during one transit, whereas the near-surface estimate gives $\tau_{\rm ST}/T_r\simeq4\times10^{-3}$. In the sign convention of Fig.~\ref{fig:schematic}, the lowest spin-energy state corresponds to $\mupar>0$ for both electrons and positrons. Sokolov--Ternov spin selection~\cite{SokolovTernov1986} therefore reinforces, rather than cancels, the SG branch imbalance and can raise $\Peff$ from an illustrative isotropic value $0.1$ toward the equilibrium magnitude $8/(5\sqrt3)\simeq0.924$. For the earlier near-surface estimate $\XiZero\simeq3.5\times10^6$, this changes $\XiEff$ from approximately $3.5\times10^5$ to $3.2\times10^6$.

This radiative regime is not included in the present simulations. Recent QED-PIC work has independently found strongly spin-polarized condensed plasmoids in radiation-reaction-dominated reconnection~\cite{Gong2025}, emphasizing that radiative spin selection and reconnection can interact in the same extreme-field domain. A pre-polarized laboratory pair beam would map to the same ordering through $\Peff\rightarrow1$ at injection. These connections motivate a future QED-radiative extension with stochastic spin flips, photon recoil, radiation reaction, and Landau-level-resolved kinetics; they do not alter the numerical conclusions of the present semiclassical parameter study.

\section{Conclusions}

The main conclusions are:
\begin{enumerate}
\item SpinPIC2D implements the complete SG--magnetization-current closure used in this semiclassical relativistic model: proper-momentum Boris pushing, magnetic BMT spin rotation, SG forcing, component-staggered magnetization deposition, filtered $\nabla\times\Mfield$, and $\Jfree+\Jm$ in Ampere's law.
\item The common Harris loading is controlled at the $10^{-3}$ level, all members of the main $\Xi$ scan remain below the one-percent total-energy-drift gate, and a fixed physical magnetization filter removes the nonphysical high-$k$ feedback branch while preserving the sheet-scale current response.
\item Increasing $\Xi$ at fixed $\gamma_{\rm tr}=2$ produces a threshold-like transition from a nearly quiescent current sheet to rapid cold-normalized global flux growth, with the same ordering in branch sorting, pressure/current restructuring, electron-frame nonideality, and accumulated flux.
\item Along the fixed-$\chi_{\rm sim}$ family, direct sorting decreases with $\gamma_{\rm tr}$ while the magnetization-current channel and matched-control flux-growth enhancement remain comparatively resilient. Holding $\Xi=1$ requires increasing $\chi_{\rm sim}$ and $J_{M,y}$; the refined $\gamma_{\rm tr}=5$, $\Xi=1$ case is energy-gated and develops a nonlinear multi-X-line state whose large normalized value is a global topology-growth calculation rather than a local reconnection rate.
\item Near-surface magnetar conditions require QED-radiative extensions. Sokolov--Ternov spin selection is expected to increase the SG-active branch participation rather than suppress it.
\end{enumerate}

\appendix

\section{Diagnostic definitions}
\label{app:diagnostics}

\subsection{Magnetic flux and normalized global flux-growth proxy}

In 2.5-D the in-plane field is represented by the reconstructed out-of-plane vector potential,
\begin{equation}
B_x=-\partial_zA_y,
\qquad
B_z=\partial_xA_y.
\end{equation}
Along the midplane,
$A_y(x,0,t)-A_y(x_0,0,t)=\int_{x_0}^{x}B_z(x',0,t)\,dx'$, up to gauge and sign.  The reported global flux per unit invariant $y$ length is
\begin{equation}
\psi(t)=\max_x A_y(x,0,t)-\min_x A_y(x,0,t).
\label{eq:psidef}
\end{equation}
For an isolated X--O pair, $\psi_{XO}=A_y(O)-A_y(X)$ and
\begin{equation}
\frac{d\psi_{XO}}{dt}=E_y(X)-E_y(O).
\end{equation}
Thus $|d\psi_{XO}/dt|\simeq|E_y(X)|$ only when the O-point field is negligible.  The conventional local Alfv\'en-normalized rate is
\begin{equation}
R_A(t)=\frac{E_y(X)-E_y(O)}{B_{\rm up}V_{A,\rm up}}
      =\frac{1}{B_{\rm up}V_{A,\rm up}}\frac{d\psi_{XO}}{dt}.
\label{eq:RA}
\end{equation}

Figures~\ref{fig:validation} and \ref{fig:relativistic} use a fully reproducible code-unit normalization based on the measured upstream reconnecting field and total pair rest-mass density,
\begin{align}
\sigma_c&=\frac{B_{\rm up}^2}{\mu_0 n_{\rm pair,up}m_ec^2},\\
\frac{V_{A,c}}{c}&=\left(\frac{\sigma_c}{1+\sigma_c}\right)^{1/2},\\
R_{\psi,c}^{\rm global}&=\frac{\dot\psi}{B_{\rm up}V_{A,c}}.
\label{eq:Rglobalcold}
\end{align}
For each snapshot, $B_{\rm up}$ and $n_{\rm pair,up}$ are the median values in the symmetric boxes $|x|\le6d_e$ and $4d_e\le|z|\le12d_e$.  The slope $\dot\psi$ is obtained by a least-squares fit to the dense flux history, collected for each run in  \texttt{energy.csv},  over $3\le t/\tausp\le7$, while $B_{\rm up}V_{A,c}$ is averaged over snapshots in the same interval.  This hybrid procedure avoids fitting a derivative from only the sparser run output-times.

A hot relativistic normalization would replace the rest-mass density by the proper enthalpy density,
\begin{equation}
\sigma_h=\frac{B_{\rm up}^2}{\mu_0w_{\rm up}},
\qquad
\frac{V_{A,h}}{c}=\left(\frac{\sigma_h}{1+\sigma_h}\right)^{1/2},
\end{equation}
with, for example, $w_{\rm up}=nmc^2+\Gamma p/(\Gamma-1)$ for an isotropic ideal closure.  Because $w_{\rm up}\ge nmc^2$, one has $V_{A,h}\le V_{A,c}$ and therefore $|R_{\psi,h}^{\rm global}|\ge|R_{\psi,c}^{\rm global}|$ for the same numerator.  The cold normalization used in the main figures is therefore the smaller (more conservative) of these two simple normalizations for a fixed numerator: it does not create the large $\gamma_{\rm tr}=5,\Xi=1$ value, and a hot normalization would increase it.  We retain Eq.~(\ref{eq:Rglobalcold}) for the main cross-run comparison because the stored pressure tensors are velocity-moment diagnostics rather than a complete covariant energy--momentum tensor, making a unique hot enthalpy reconstruction model dependent.

Once multiple X and O points form, $\dot\psi$ includes island growth, merger, and motion of the global extrema.  In that regime $R_{\psi,c}^{\rm global}$ is not a unique local $R_A$ and values larger than unity do not imply a super-Alfv\'enic plasma inflow.  Negative control values likewise indicate relaxation of the imposed peak-to-peak seed during the fit interval, not reverse reconnection.  For relativistic-family comparisons we therefore also report the matched-control enhancement
\begin{equation}
\Delta R_{\psi,c}=R_{\psi,c}^{\rm full}-R_{\psi,c}^{\rm ctrl}.
\end{equation}
The unnormalized early slope in Fig.~\ref{fig:xiscan} is evaluated over the same time interval and is used only for matched-run ordering.

\subsection{Sorting, current, topology, percentiles, and energy}

The branch asymmetry is
\begin{equation}
A_\mu(\mathbf x,t)=\frac{n_{\mu+}-n_{\mu-}}{n_{\mu+}+n_{\mu-}},
\end{equation}
where the branches are classified by the sign of $\mupar$. The scalar sorting metric is the mean $\avg{|A_\mu|}$ over the stated region. The Harris current normalization in Fig.~\ref{fig:relativistic_maps} is $J_{H,0}=B_0/\lambda$ in the code normalization. The magnetic response is $\Delta A_y(\mathbf x,t)=A_y(\mathbf x,t)-A_y(\mathbf x,0)$. ``p99'' denotes the 99th percentile of the absolute finite cell values in the indicated region; the Fig.~\ref{fig:xiscan} spatial metrics use $|x|\le1.5\de$ and $|z|\le1.0\de$ at $t\simeq7\tausp$.

The operational energy metric is
\begin{equation}
\delta_E(t)=100\left|\frac{E_{\rm code}(t)-E_{\rm code}(0)}{E_{\rm code}(0)}\right|,
\end{equation}
with acceptance criterion $\max_t\delta_E<1\%$. The Supplemental Material gives the separate particle, electromagnetic, SG, and magnetization-current work ledgers.

\subsection{High-$k$ and pressure-balance diagnostics}

For a two-dimensional current map, the high-$k$ fraction in Fig.~\ref{fig:highk} is
\begin{equation}
f_{>k_c}=\frac{\sum_{|\mathbf k|>k_c}|\widetilde J_{M,y}(\mathbf k)|^2}{\sum_{\mathbf k}|\widetilde J_{M,y}(\mathbf k)|^2},
\end{equation}
with the same cutoff $k_c$ for every smoothing case. The Harris pressure residual in Fig.~\ref{fig:harris} is $100(P_H/P_{\rm up}-1)$, where $P_H$ uses the pair thermal pressure prescribed by the loading plus the unperturbed Harris magnetic pressure $B_H^2/2$ in code units.

\section{Generalized-Ohm's law diagnostic terms}
\label{app:ohm}

The electron-frame nonideal field is $E'_y=E_y+(\mathbf u_e\times\mathbf B)_y$. The code evaluates
\begin{align}
\mathcal P_e&=\frac{(\nabla\cdot\mathbf P_e)_y}{q_en_e},\\
\mathcal I_e&=\frac{m_e}{q_e}\left(\partial_tu_{e,y}+\mathbf u_e\cdot\nabla u_{e,y}\right),\\
\mathcal S_e&=-\frac{F_{{\rm SG},e,y}}{q_en_e},\\
R_e&=E'_y-\mathcal P_e-\mathcal I_e-\mathcal S_e.
\end{align}
For the present $\partial_y=0$ geometry, $F_{{\rm SG},e,y}=0$. The p99 and RMS amplitudes in Fig.~S5 are evaluated in the same $|x|\le1.5\de$, $|z|\le1.0\de$ X-line box used for Fig.~\ref{fig:xiscan}. The structured residual is an empirical closure diagnostic and is not identified with a constant collisional resistivity.

\section*{Data Availability}
The archived simulation products needed to reproduce the figures in
this article and its Supplemental Material, including selected original
HDF5 snapshots, compact time histories, derived CSV tables, run
settings, logs, figure-generation and analysis scripts, manifests, and
SHA-256 checksums, are available from Figshare at
\href{https://doi.org/10.6084/m9.figshare.33007145}
The SpinPIC2D source code is not included in this release and will be
made available separately subject to institutional approval.

\begin{acknowledgments}
The author thanks Brandon K. Russell and Hantao Ji of the Princeton
Plasma Physics Laboratory for discussions that motivated the radiative
spin-selection comparison and the connection to spin-polarized pair
beams. The theory and concept of applying the Stern--Gerlach force to magnetic
reconnection was first developed by the author in 2015. Author also acknowledges Embry-Riddle's
start-up-research fund and support for Claude subscription. Claude Code and Claude Sonnet (Anthropic), together with ChatGPT (OpenAI), were used as
AI-assisted programming tools during the development and debugging of
SpinPIC2D and during the development of Python analysis, diagnostic,
and figure-generation scripts. All AI-assisted code was reviewed,
modified as needed, and tested and validated by the author. The author
takes full responsibility for the numerical implementation, scientific
interpretation, and content of the manuscript.
\end{acknowledgments}

\clearpage
\onecolumngrid
\newcount\smp \smp=1
\loop
\noindent\makebox[\textwidth]{\includegraphics[page=\the\smp,height=0.993\textheight,keepaspectratio]{Nykyri_PRE_SM_Jul16_FINAL_submission.pdf}}\clearpage
\advance\smp 1
\ifnum\smp<20
\repeat
\end{document}